\documentclass[11pt]{article}

\usepackage[english]{babel}

\usepackage[a4paper,top=2cm,bottom=2cm,left=2.5cm,right=2.5cm,marginparwidth=1.75cm]{geometry}
\usepackage{setspace}
\usepackage{amsmath,amssymb}
\usepackage{graphicx}
\usepackage{authblk}
\usepackage[colorlinks=true, allcolors=blue]{hyperref}
\usepackage[utf8]{inputenc}
\usepackage[T1]{fontenc}
\usepackage[numbers]{natbib}
\usepackage{hyperref}
\usepackage{titlesec}
\usepackage{caption}
\usepackage{setspace}
\usepackage{multicol}
\usepackage{ragged2e} 
\usepackage[dvipsnames]{xcolor} 
\usepackage{tcolorbox} % Paquete para la cajita de color del abstract 

\title{Why the Northern Hemisphere Needs a 30-40m Telescope and the Science at Stake: from Interstellar Visitors to Planetary Defence}
\author[1,2]{J. de León}
\author[3]{N. Pinilla-Alonso}

\author[4]{P. Tanga}
\author[5]{D. Souami}
\author[6]{Z. Gray}

\author[7]{A. Alvarez-Candal}
\author[4]{B. Carry}
\author[8]{R. de la Fuente Marcos}
\author[9]{A. Delsanti}
\author[10]{F. La Forgia}
\author[11]{A. Migliorini}
\author[12]{T. Müller}
\author[6]{A. Penttilä}
\author[13,14]{M. Popescu}
\author[15]{C. Snodgrass}

\author[16]{D. Oszkiewicz}
\author[15]{C. Opitom}
\author[17]{A. Campo-Bagatin}
\author[1,2]{J. Licandro}
\author[18]{R. Hueso}
\author[10]{M. Lazzarin}
\author[5]{S. Fornasier}

\author[19]{R. Brunetto}
\author[3]{J. A. de Abol Brasón}
\author[3]{J. de Cos Juez}
\author[6]{J. DeMartini}
\author[15]{A. Donaldson}
\author[6]{R. Dorsey}
\author[7]{R. Duffard}
\author[3]{J. Fernández Díaz}
\author[3]{F. García de Leániz}
\author[3]{R. Hevia Díaz}
\author[7]{J. M. Gómez-Limón}
\author[9]{O. Groussin}
\author[3]{S. Iglesias Álvarez}
\author[20]{T. Kohout}
\author[21]{M. Kretlow}
\author[1,2]{T. Le Pivert-Jolivet}
\author[3]{M. Montero-Vega}
\author[7]{N. Morales}
\author[6]{K. Muinonen}
\author[7]{J. L. Ortiz}
\author[1,2]{G. P. Prodan}
\author[7]{J. L. Rizos}
\author[15]{J. E. Robinson}
\author[3]{S. Rodríguez Cabo}
\author[3]{J. Rodríguez Rodríguez}
\author[15]{A. Ro\.{z}ek}
\author[7]{P. Santos-Sanz}
\author[1,2]{E. Tatsumi}
\author[4]{F. Tinaut-Ruano}
\author[1,2]{E. Villaver}

\affil[1]{Instituto de Astrofísica de Canarias - IAC, Tenerife, Spain}
\affil[2]{Departamento de Astrofísica, Universidad de La Laguna, Tenerife, Spain}
\affil[3]{Instituto de Ciencias y Tecnologías Espaciales de Asturias - ICTEA, Universidad de Oviedo, Asturias, Spain}
\affil[4]{Université Côte d’Azur, Observatoire Côte d’Azur, CNRS, Laboratoire Lagrange, Nice, France}
\affil[5]{LIRA, Observatoire de Paris, CNRS, Université PSL, Université Paris Cité, Sorbonne Université, Meudon, France}
\affil[6]{Department of Physics, University of Helsinki, Finland}
\affil[7]{Instituto de Astrofísica de Andalucía - CSIC, Granada, Spain}
\affil[8]{Universidad Complutense de Madrid, Ciudad Universitaria, Madrid, Spain}
\affil[9]{Aix Marseille University, CNRS, CNES, Laboratoire d’Astrophysique de Marseille - Marseille, France}
\affil[10]{Dipartimento di Fisica e Astronomia, Università di Padova, Padova, Italy}

\affil[11]{Institute for Space Astrophysics and Planetology  - IAPS-INAF, Rome, Italy}
\affil[12]{Department of High-Energy Astrophysis,  Max-Planck-Institut für Extraterrestrische Physik, Giessenbachstrasse, Garching, Germany}
\affil[13]{Institute of Space Science - INFLPR subsidiary, Măgurele, Romania}
\affil[14]{University of Craiova, Craiova, Romania}
\affil[15]{Institute for Astronomy, University of Edinburgh, Royal Observatory, Edinburgh, UK}
\affil[16]{Institute Astronomical Observatory, Faculty of Physics and Astronomy, Adam Mickiewicz University, Poznań, Poland}
\affil[17]{Instituto de Física Aplicada a las Ciencias y las Tecnologías, Universidad de Alicante, Alicante, Spain}
\affil[18]{Escuela de Ingeniería de Bilbao, Universidad del País Vasco UPV/EHU, Bilbao, Spain}
\affil[19]{Université París-Saclay, CNRS, IAS, Orsay, France}
\affil[20]{Department of Electronics and Nanoengineering, Aalto University, Finland}
\affil[21]{Deutsches Zentrum für Astrophysik (DZA), Görlitz, Germany}

\begin{document}
\maketitle

\newpage

\begin{tcolorbox}[colback=RoyalBlue!5!white,colframe=black!75!black, width=\textwidth]
\justifying
{\noindent Small Solar system Objects (SSOs) preserve the physical, chemical, and dynamical signatures of the Sun's protoplanetary disk. Upcoming surveys will discover vast numbers of new objects, yet their scientific value will depend on follow-up observations requiring far greater sensitivity and resolution than those currently available. A 30-m class telescope like the Extremely Large Telescope (ELT) will be transformative, but its Southern location leaves significant regions of the sky poorly covered or even non accessible.{\bf A Northern 30-40m telescope is therefore essential to achieve full-sky coverage and fully exploit the small body discoveries of the 2030–2050 era}, in particular for targets of opportunity or unexpected discoveries, like those of interstellar objects and potentially hazardous asteroids, as well as for distant trans-Neptunian objects and space mission targets.}
\end{tcolorbox}
\vspace{-10pt}
\section{Introduction and Motivation}
\vspace{-10pt}
Forthcoming advancements in survey facilities, most notably the Vera C. Rubin Observatory but also the NEO Surveyor mission \cite{2023PSJ.....4..224M}, will vastly expand the known inventory of small bodies. However, without a 30-m class facility like the ELT, the scientific return from these discoveries will be limited, and key science questions, including the origin of water in the inner Solar System and the search for the origin of life, the planetary differentiation processes and the collisional history of SSOs, or the physical drivers of cometary activity, will remain only partially explored. Looking toward the 2040s, while the ELT will transform many areas of planetary science, its Southern location will leave substantial portions of the ecliptic and northern sky inaccessible or only marginally observable, potentially leading to missed chances for discovering target-of-opportunity objects, such as interstellar objects and Earth's virtual impactors. A northern 30-m telescope is therefore not redundant but rather highly complementary and necessary, and together, the northern and southern facilities would enable full-sky access to the most interesting small bodies, ensuring that the global astronomical community fully exploits the scientific potential of the upcoming era, as evidenced by the achievements of recent years.
\vspace{-10pt}
\section{The Science Challenge}
\vspace{-10pt}
A 30-40m telescope facility in the Northern hemisphere presents unique opportunities for SSOs studies over long time scales: 1) Northern hemisphere surveys are less affected by the crowdedness of the densest galactic areas (including the bulge), that are present in the Southern counterpart. Deep searches for faint objects are thus less biased and can provide more uniform coverage. This will improve the search for faint trans-Neptunian objects (TNOs) requiring uniform coverage to dynamically constrain the existence and position of potential distant massive perturbers (e.g. Planet X); 2) For the next decades, giant planets will have positive declinations, thus favoring observations from the North. That’s the case for Saturn (until 2040) and Uranus (until 2050). Neptune will be favored for the next $\sim$80 years, and so will its Trojan clouds, in particular the L4 \cite{murtagh2025predictionslsstsolaryield}). Proper characterization of primordial Trojans is essential to constrain the evolution of the Solar System, and  for understanding the formation and evolution of planetary systems.

We identify several scientific questions that will be adversely impacted due to a combination of a loss of sky coverage and opportunities, if a northern 30-40m class telescope in the 2040s is not available. 

\vspace{-10pt}
\subsection{Interstellar Objects (ISOs)}
The importance of having the above described capabilities has grown even more critical with the recent discovery of interstellar objects (ISOs) entering the Solar System. The first detections, 1I/‘Oumuamua in 2017 \cite{2018NatAs...2..133F,2019NatAs...3..594O},  2I/Borisov in 2019 \cite{2020MNRAS.495.2053D}, and 3I/ATLAS in 2025 \cite{2025A&A...700L...9D,2025ApJ...991L..43C,2025ApJ...992L...9Y}, revealed that material from other planetary systems regularly traverses interplanetary space, offering unprecedented opportunities to directly investigate the building blocks of planetary systems beyond the Sun. Discovered ISOs had {\bf short observability windows} and rapid sky motions (see Fig.1). Characterizing their physical and chemical properties requires sensitive, high-resolution spectroscopy and imaging on timescales of hours to days, before they get too faint. Consequently, a 30-40m class facility will allow for a complete characterization at inbound (and outbound) trajectories at distances before the onset of cometary activity or significant modification of its surface by solar heating, resulting in {\bf unique, transformative science}. The spatial distribution of incoming ISOs is expected to be essentially isotropic, reflecting the random velocities of planetesimals ejected from other stellar systems throughout the Galaxy. Interestingly, recent models place the most likely radiants towards the Northern hemisphere \cite{2025ApJ...990L..30H}, in the direction of the Solar Apex ($\sim$18h30, +30deg N). In the absence of a Northern 30-40m class telescope we will be disadvantaged in the characterization of high inclination objects: those entering from the South would be discovered and tracked initially, but then potentially lost (likely after perihelion passage), while those approaching from the North would not be detected from the South until late (or too late), and possibly in unfavorable observing conditions. Both situations must be avoided to fully characterize these objects during their inbound and outbound paths and to compare extrasolar planetesimals with native SSOs. 

\begin{figure}
\centering
\begin{minipage}{0.5\linewidth}
    \includegraphics[width=\linewidth]{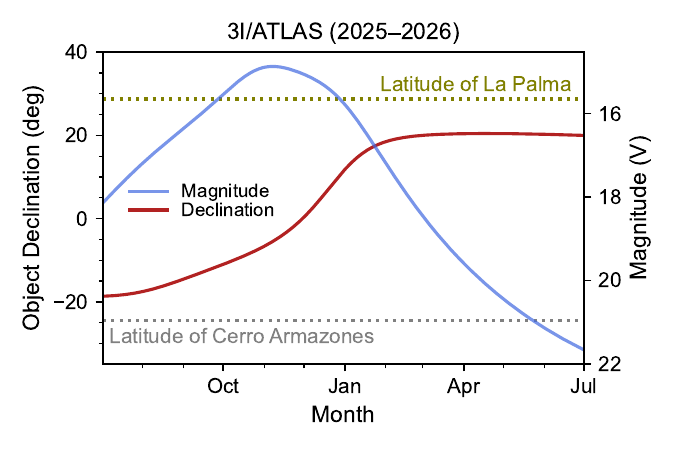}
\end{minipage}
\hfill
\begin{minipage}{0.45\linewidth}
\caption{Declination and apparent visual magnitude of interstellar object 3I/ATLAS between July 1, 2025 and July 1, 2026. The dotted horizontal lines indicate the latitude of Cerro Armazones (location of ELT) and La Palma (as reference for a Northern location).}
\label{fig:frog}
\end{minipage}
\end{figure}

\vspace{-10pt}
\subsection{Distant trans-Neptunian objects}
Distant or extreme trans-Neptunian objects (ETNOs) are defined as bodies with perihelia beyond $\sim$50 au and semi-major axes exceeding $\sim$150 au. A 30-40m telescope is indispensable for detailed physical and compositional studies of ETNOs: these objects are very faint ($V >$ 25), often at magnitudes beyond the reach of current 8–10m class facilities, and mid-resolution spectroscopy and imaging are required to determine their surface composition, detect binary companions, and measure rotation periods. In the last years, JWST spectroscopy from 2 to 5 $\mu$m has proven successful to study compositional diversity, from Neptune Trojans to ETNOs \cite[e.g][]{2025NatAs...9..230P,2025PSJ.....6..154M}. However, at the time of this writing, less than 6 out of the more than 40 currently known ETNOS with reliable orbital determinations have any physical or compositional information. Such observations will provide much needed insight on the formation conditions in the outermost regions of the Solar System, the collisional history of the trans-Neptunian belt, and the migration history of the giant planets. 

Detailed orbital characterization, requiring precise astrometry over multiple oppositions, is essential to constrain the orbits of these objects, detect subtle clustering in their orbital elements, and distinguish between random distributions and gravitational sculpting. Gaps in the orbital coverage, that affect more strongly southern declinations due to locally extreme stellar density, would prevent high-precision modeling of their dynamical evolution. Failure to build a northern 30-m class facility would mean that a substantial fraction of distant TNOs discovered by upcoming surveys, including those most dynamically sensitive to the hypothesized Planet X \cite{2016AJ....151...22B}, would remain poorly studied or non characterized, leaving a significant portion of the outer Solar System effectively inaccessible, and therefore, limiting our knowledge of the collisional history of the trans-Neptunian population, and the migration history of the giant planets.

\vspace{-10pt}
\subsection{Space mission support and coordination}
A 30-40m telescope in the Northern hemisphere may be a critical asset for supporting future spacecraft missions to small bodies, particularly for targets which are primarily or exclusively observable from Northern latitudes in the years preceding encounter. Ground-based observations are a key element of successful mission planning, both for maximizing scientific return (early characterization of physical properties) and for spacecraft control and engineering purposes (i.e., reduction of ephemeris uncertainties). Also important are the rapid-response observations associated with active mission events, impacts, flybys, or other transient phenomena, which often occur at specific geometries relative to Earth. Two past examples are 1) the NASA Deep Impact mission, which struck 9P/Tempel 1 in July 2005, excavating subsurface material and producing a rapidly evolving ejecta plume. Although an extensive worldwide campaign was organized \cite{2005Sci...310..265M}, only a small number of ground-based observatories were able to observe the moment of impact, in particular those stationed in Hawaii \cite{2007Icar..191S.371B,2007Icar..191S.424F}; 2) the DART impact to Dimorphos in September 22, also excavating a rapid evolving plume. In contrast to Deep Impact, no major ground-based facility captured the moment of impact; instead, the first VLT-class observations began hours after \cite{2024PSJ.....5...18G,2023A&A...671L..11O}.
Earth-based accessibility, especially in the first minutes post-event, is strongly latitude and longitude dependent. Although we cannot predict which objects or mission events will be most crucial in the coming decades, experience shows that rapid follow-up ground-based observations maximize scientific output of space missions. A Northern 30-40m facility would provide flexibility and capability needed to respond to these developments.

\vspace{-10pt}
\subsection{Earth virtual impactors: Planetary Defence}
Beyond advancing fundamental science, a Northern 30-40m class telescope would play a critical role in planetary defence. At the time of this writing, the total number of near-Earth asteroids (NEAs) is slightly over 40,000. Of those, asteroids having a diameter $D >$ 140 m and orbits that bring them at distances to Earth $<$ 0.05 au, are classified as potentially hazardous (PHAs). While we have identified all  NEAs with $D >$ 1 km, our current estimates indicate that we have only identified $\sim$50\% of NEAs in the size range between 140 m and 1 km (small asteroids). Therefore, a substantial fraction of PHAs and virtual impactors (VIs), objects whose orbits carry a non-zero probability of Earth collision, {\bf remain small, faint, or located at high northern declinations}. Rapid, high-sensitivity follow-up observations are essential to refine orbital parameters, assess impact probabilities, and model physical properties such as rotation, shape, composition, and interior structure \cite{2023A&A...676A.126P}. The combination of high angular resolution (to reduce ephemeris uncertainties) and spectral sensitivity provided by a 30-40m telescope would allow detailed compositional and surface characterization of PHAs and VIs, enabling better predictions of their mechanical behavior and potential response to mitigation efforts. Determining whether a small body is a rubble pile, a monolithic rock, or volatile-rich can inform the design of mitigation strategies, including deflection missions or kinetic impactor strategies. This was the case of the NASA DART mission and its target, asteroid (65803) Didymos \cite{2023Natur.616..448T,2023PSJ.....4..214R}. 
In this sense, Northern facilities do not merely complement Southern surveys; they are an {\bf indispensable component of a global planetary defence infrastructure} capable of providing full-sky coverage and rapid response in the event of a near-Earth threat. A good example was the case of asteroid 2024 YR4, that activated for the first time in human history the UN Planetary Defence Protocol, and needed for large telescopes at all latitudes. In addition, both the NEO Surveyor and, in a more distant future, the ESA NEOMIR telescope \cite{2025epsc.conf.1329C}, will also require the support of large, ground-based facilities in the North. Without access to a 30-40m  Northern-aperture facility, critical warnings could be delayed, leaving limited time for mitigation strategies or deflection campaigns in the event of a threatening object. This is currently of major concern for space agencies such as NASA and ESA, and for the European Commission.
\vspace{-10pt}
\section{Capability Requirements}
\vspace{-10pt}
We include here some technical requirements needed for the scientific cases described above: 1) Wide field visible imaging for NEAs/PHAs/VIs and ISOs characterization and astrometry/recovery, as well as for comet characterization; 2) Small field fast imaging with absolute sub-ms time tagging capabilities for local high-precision astrometry of TNOs and for occultations; 3) High-contrast and high-resolution imaging (AO) for large TNOs; 4) Low-mid resolution from near-UV (0.3 $\mu$m) up to mid-infrared (2-5 $\mu$m and 10 $\mu$m) spectroscopy for compositional characterization of NEAs/PHAs/VIs, TNOs, gas emission in comets and ISOs; 5) Spectro-polarimetric capabilities, in particular for ISOs, as their phase angle varies quickly; 6) Timing accuracy and tracking modes; 7) IFU + AO (spaxel of few mas) + mid-resolution, to characterize extended objects (disk-resolved large objects) + coma in active objects + resolving close multiple systems.

\begin{multicols}{2}
\renewcommand{\bibfont}{\footnotesize}  % Tamaño pequeño
\bibliographystyle{unsrtnat}     % o el estilo que prefieras
\bibliography{References}

@ARTICLE{2018NatAs...2..133F,
       author = {{Fitzsimmons} et al.},
      journal = {Nat. Astron.},
         year = 2018,
       volume = {2},
        pages = {133-137},
       adsurl = {https://ui.adsabs.harvard.edu/abs/2018NatAs...2..133F}
}

@ARTICLE{2019NatAs...3..594O,
       author = {{'Oumuamua ISSI Team} et al.},
      journal = {Nat. Astron.},
         year = 2019,
       volume = {3},
        pages = {594-602},
       adsurl = {https://ui.adsabs.harvard.edu/abs/2019NatAs...3..594O}
}

@ARTICLE{2020MNRAS.495.2053D,
       author = {{de Le{\'o}n} et al.},
      journal = {MNRAS},
         year = 2020,
       volume = {495},
       number = {2},
        pages = {2053-2062},
       adsurl = {https://ui.adsabs.harvard.edu/abs/2020MNRAS.495.2053D}
}

@ARTICLE{2025A&A...700L...9D,
       author = {{de la Fuente Marcos} et al.},
      journal = {A\&A},
         year = 2025,
       volume = {700},
          eid = {L9},
        pages = {L9},
       adsurl = {https://ui.adsabs.harvard.edu/abs/2025A&A...700L...9D}
}

@ARTICLE{2025ApJ...991L..43C,
       author = {{Cordiner} et al.},
      journal = {ApJL},
         year = 2025,
       volume = {991},
       number = {2},
          eid = {L43},
        pages = {L43},
       adsurl = {https://ui.adsabs.harvard.edu/abs/2025ApJ...991L..43C}
}

@ARTICLE{2025ApJ...992L...9Y,
       author = {{Yang} et al.},
      journal = {ApJL},
         year = 2025,
       volume = {992},
       number = {1},
          eid = {L9},
        pages = {L9},
       adsurl = {https://ui.adsabs.harvard.edu/abs/2025ApJ...992L...9Y}
}

@ARTICLE{2023Natur.616..448T,
       author = {{Thomas} et al.},
      journal = {Nature},
         year = 2023,
       volume = {616},
       number = {7957},
        pages = {448-451},
       adsurl = {https://ui.adsabs.harvard.edu/abs/2023Natur.616..448T}
}

@ARTICLE{2023PSJ.....4..214R,
       author = {{Rivkin} et al.},
      journal = {PSJ},
         year = 2023,
       volume = {4},
       number = {11},
          eid = {214},
        pages = {214},
       adsurl = {https://ui.adsabs.harvard.edu/abs/2023PSJ.....4..214R}
}

@ARTICLE{2016AJ....151...22B,
       author = {{Batygin} \& {Brown}},
      journal = {AJ},
         year = 2016,
       volume = {151},
       number = {2},
          eid = {22},
        pages = {22},
       adsurl = {https://ui.adsabs.harvard.edu/abs/2016AJ....151...22B}
}

@ARTICLE{murtagh2025predictionslsstsolaryield,
      author={{Murtagh} et al.},
      journal={arXiv},
      year={2025},
      volume={2512},
      pages = {03892}
}

@ARTICLE{2025ApJ...990L..30H,
       author = {{Hopkins} et al.},
      journal = {ApJL},
         year = 2025,
       volume = {990},
       number = {2},
          eid = {L30},
        pages = {L30},
       adsurl = {https://ui.adsabs.harvard.edu/abs/2025ApJ...990L..30H}
}

@ARTICLE{2005Sci...310..265M,
       author = {{Meech} et al.},
      journal = {Science},
         year = 2005,
       volume = {310},
       number = {5746},
        pages = {265-269},
       adsurl = {https://ui.adsabs.harvard.edu/abs/2005Sci...310..265M}
}

@ARTICLE{2007Icar..191S.371B,
       author = {{Barber} et al.},
      journal = {Icarus},
         year = 2007,
       volume = {191},
       number = {2},
        pages = {371-380},
       adsurl = {https://ui.adsabs.harvard.edu/abs/2007Icar..191S.371B}
}

@ARTICLE{2007Icar..191S.424F,
       author = {{Fern{\'a}ndez} et al.},
      journal = {Icarus},
         year = 2007,
       volume = {191},
       number = {2},
        pages = {424-431},
       adsurl = {https://ui.adsabs.harvard.edu/abs/2007Icar..191S.424F}
}

@ARTICLE{2024PSJ.....5...18G,
       author = {{Gray} et al.},
      journal = {PSJ},
         year = 2024,
       volume = {5},
       number = {1},
          eid = {18},
        pages = {18},
       adsurl = {https://ui.adsabs.harvard.edu/abs/2024PSJ.....5...18G}
}

@ARTICLE{2023A&A...671L..11O,
       author = {{Opitom} et al.},
      journal = {A\&A},
         year = 2023,
       volume = {671},
          eid = {L11},
        pages = {L11},
       adsurl = {https://ui.adsabs.harvard.edu/abs/2023A&A...671L..11O}
}

@ARTICLE{2025PSJ.....6..154M,
       author = {{Markwardt} et al.},
      journal = {PSJ},
         year = 2025,
       volume = {6},
       number = {7},
          eid = {154},
        pages = {154},
       adsurl = {https://ui.adsabs.harvard.edu/abs/2025PSJ.....6..154M}
}

@ARTICLE{2025NatAs...9..230P,
       author = {{Pinilla-Alonso} et al.},
      journal = {Nat. Astron.},
         year = 2025,
       volume = {9},
        pages = {230-244},
       adsurl = {https://ui.adsabs.harvard.edu/abs/2025NatAs...9..230P}
}

@ARTICLE{2023PSJ.....4..224M,
       author = {{Mainzer} et al.},
      journal = {PSJ},
         year = 2023,
       volume = {4},
       number = {12},
          eid = {224},
        pages = {224},
       adsurl = {https://ui.adsabs.harvard.edu/abs/2023PSJ.....4..224M}
}

@ARTICLE{2025epsc.conf.1329C,
       author = {{Conversi} et al.},
      journal = {EPSC-DPS 2025},
         year = 2025,
       volume = {18},
        pages = {1329},
       adsurl = {https://ui.adsabs.harvard.edu/abs/2025epsc.conf.1329C}
}

@ARTICLE{2023A&A...676A.126P,
       author = {{Popescu} et al.},
      journal = {A\&A},
         year = 2023,
       volume = {676},
          eid = {A126},
        pages = {A126},
      adsurl = {https://ui.adsabs.harvard.edu/abs/2023A&A...676A.126P}
}
\end{multicols}

\end{document}